\documentclass[twocolumn,aps,showpacs,showkeys]{revtex4}
\usepackage{epsfig,amsmath,amssymb}
\usepackage{amsmath}

\begin{document}

\title{Low-temperature thermodynamics of spin-1/2 orthogonal-dimer chain
with Ising and Heisenberg interactions}
\author{T. Verkholyak$^{a,b}$,
        J. Stre\v{c}ka$^{b}$}
\affiliation{$^a$Institute for Condensed Matter Physics,
             National Academy of Sciences of Ukraine,
             1 Svientsitskii Street, L'viv-11, 79011, Ukraine\\
             $^b$Department of Theoretical Physics and Astrophysics,
             Institute of Physics, P. J. \v{S}af\'{a}rik University,
             Park Angelinum 9, 040 01 Ko\v{s}ice, Slovak Republic}
\date{\today}

\begin{abstract}
We consider an exactly solvable version of the quantum spin-1/2 orthogonal-dimer chain
with the Heisenberg intra-dimer and Ising inter-dimer couplings.
The investigated quantum spin system exhibits at zero
temperature fractional plateaux at 1/4 and 1/2 of the saturation magnetization
and it has a highly degenerate ground state at critical fields where the
magnetization jumps.
We study the field dependence of the specific heat at low temperature.
The lattice-gas description is formulated in a vicinity of critical fields
to explain the low-temperature behaviour of specific heat.
\end{abstract}



\maketitle

The orthogonal-dimer chain \cite{richter1998} constitutes a one-dimensional
counterpart of the well-known Shastry-Sutherland lattice. The spin models on
the latter lattice are intensively studied in connection with a number of fractional
magnetization plateaux experimentally observed in SrCu$_2$(BO$_3$)$_2$, RB$_4$
(R denotes the rare-earth element) and other magnetic compounds with competing
interactions \cite{takigawa2011}.

In present paper we study a special version of the orthogonal-dimer chain with the Ising and Heisenberg interactions,
which has been solved recently \cite{paulinelli2013,odc}.
The field dependence of the specific heat is explored at low temperatures
and the lattice-gas model is used to describe the low-temperature thermodynamics near critical fields.

The quantum spin-1/2 Heisenberg-Ising orthogonal-dimer chain (see Fig.1 in Ref.\onlinecite{odc})
considered here is described by the following Hamiltonian:
\begin{eqnarray}
\label{ham_gen}
H&{=}&\sum_{i=1}^N H_i,
\\
H_{2i+1}&{=}&
J_1[(s_{1,2i}^z{+}s_{2,2i}^z)s_{1,2i+1}^z{+}s_{2,2i+1}^z(s_{1,2i}^z{+}s_{2,2i}^z)]
\nonumber\\&&
+J({\mathbf s}_{1,2i+1}\cdot{\mathbf s}_{2,2i+1})
{-}h(s_{1,2i+1}^z{+}s_{2,2i+1}^z),
\nonumber\\
H_{2i}&{=}&
J({\mathbf s}_{1,2i}\cdot{\mathbf s}_{2,2i})-h(s_{1,2i}^z+s_{2,2i}^z),
\nonumber
\end{eqnarray}
where
$s_{l,i}^\alpha$ denotes spatial projections ($\alpha=x,y,z$) of the spin-$\frac{1}{2}$ operator,
$J$ is the Heisenberg intra-dimer interaction between spins on vertical and horizontal bonds,
and $J_1$ is the Ising inter-dimer interaction between spins from different bonds,
$h$ is the external magnetic field.
Since $z$-component of the total spin on the Heisenberg
dimers becomes a conserved quantity, the model under investigation
can be presented as a classical spin chain and solved exactly
\cite{paulinelli2013,odc}.

For easy reference, we mention here briefly the results for the ground state obtained in Ref. \cite{odc}.
The zero-field ground state for $J_1<\sqrt{2}J$ is in the singlet-dimer (SD) phase, which is defined
by the product of singlet dimers residing on all horizontal and vertical bonds.
The zero-field ground state changes for $J_1>\sqrt{2}J$ to the modulated antiferromagnetic (MAF) phase,
which is characterized by antiferromagnetic ordering of completely polarized vertical bonds.
When the external field is switched on, both zero-field ground states are first changed to the modulated ferrimagnetic (MFI) phase
characterized by staggering of the polarized and singlet states on vertical bonds (1/4 plateau) and subsequently
to the staggered bond (SB) phase characterized by staggering of the same states on both types of bonds (1/2 plateau).
Finally, all spins are aligned in the field direction for strong enough magnetic fields within the completely ordered ferromagnetic (FM) state.
It was also observed that the ground state for critical fields is macroscopically degenerate.

The macroscopic degeneracy is of course lifted when the field deviates from the critical value.
If the field is close to its critical value, the low-energy states can be considered within an effective lattice-gas model.
Such a treatment was used for localized magnon states in frustrated magnets \cite{zhitomirsky2004,derzhko2006}.
If one considers in our model a phase boundary between SD and MFI phases, both phases as well as their mixture have the same energy.
If an empty site on some fictitious lattice is assigned to the singlet state on a vertical bond and an occupied site to the polarized spin up state,
we can present the problem as a lattice gas with an infinitely strong repulsion between particles on neighbouring sites.
The degeneracy is lifted in the vicinity of critical field, since the change of state from a singlet to polarized
on one vertical bond (i.e. deposition of particle) costs the energy $-(h-h_c^{(1/4)})$, where $h_c^{(1/4)}=-\sqrt{J^2+J_1^2}+\frac{J(3+\Delta)}{2}$.
In general, such an effective lattice-gas model can be presented by the following Hamiltonian:
\begin{eqnarray}
\label{ham_dimer}
H^{SD-MFI}=\sum_{i=1}^{N/2}Vn_i n_{i+1} - (h-h_c^{(1/4)})n_i,
\end{eqnarray}
where $n_i=0,1$ is the occupation number and $V\to\infty$ describes an infinitely strong repulsion.
Here, the sum extends over the vertical bonds only. The same holds also in the vicinity of the SB-FM border,
where the magnetization jumps from 1/2 to the saturated value at $h_c^{(1)}=\frac{J(1+\Delta)}{2} + J_1$.
The only difference is that the configurations of all bonds should be taken into account in the latter case
and that the notation for empty and occupied sites should be inverted.

\begin{figure}[t]
 \begin{center}
   \epsfig{file=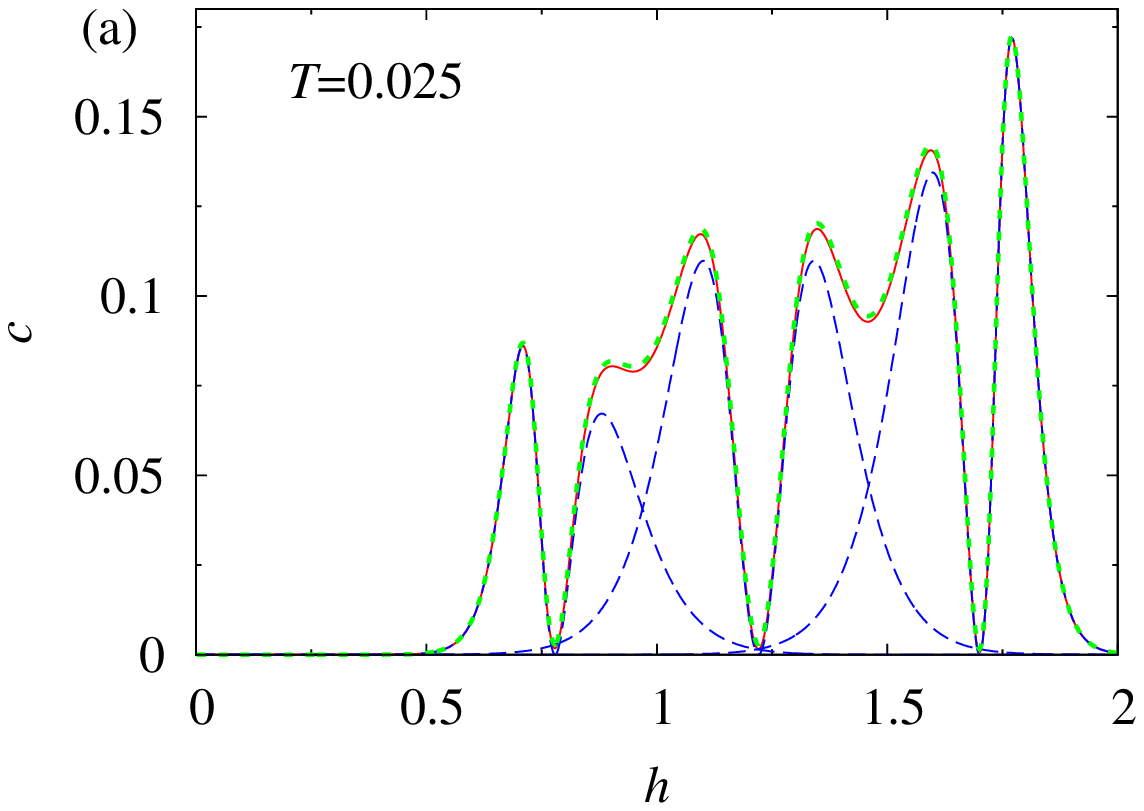, width=0.45\textwidth}
   \epsfig{file=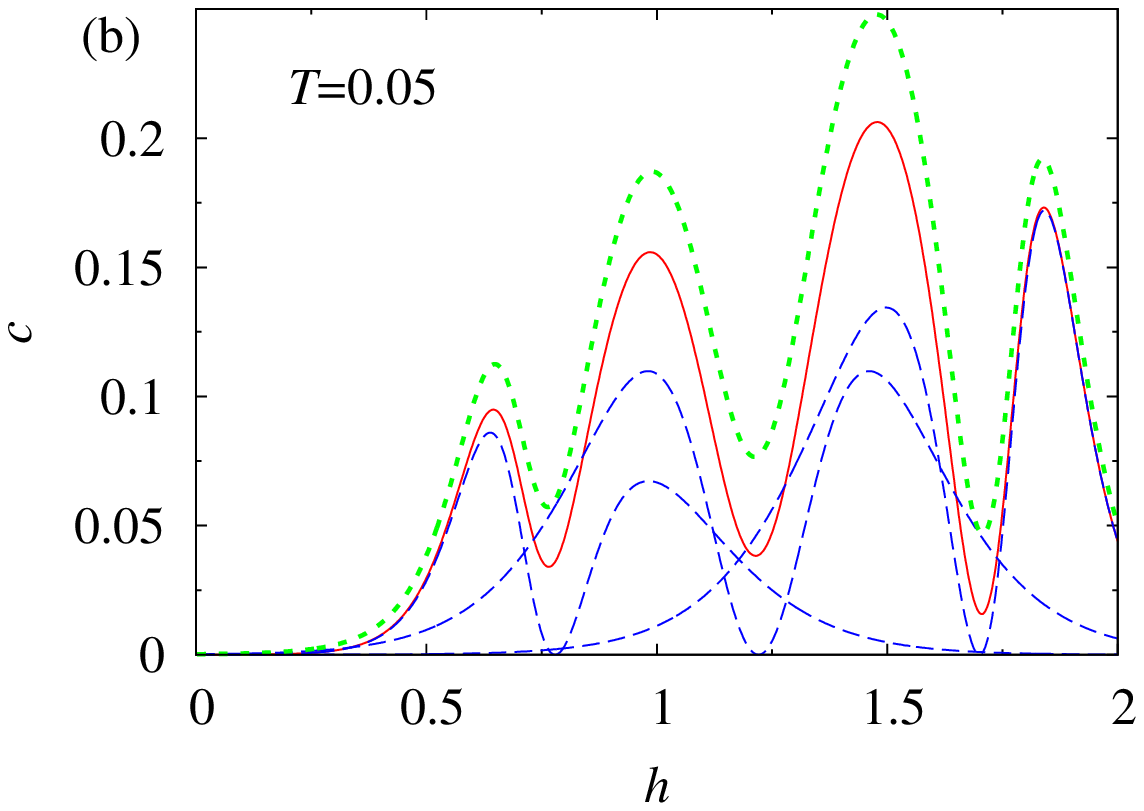, width=0.45\textwidth}
\end{center}
\vspace*{-20pt}
\caption{Low-temperature specific heat vs. magnetic field $h$:
$J=1$, $J_1=0.7$ $T=0.025$ (a),  $T=0.05$ (b). Solid (red) curves show the exact result,
dashed (blue) curves show the result of the lattice-gas model near each critical point,
dotted (green) curves show the sum of all lattice-gas models' contributions.}
\label{fig_heat}
\end{figure}

The similar procedure is valid also close to the MFA-SB boundary ($h_c^{(1/2)}=\sqrt{J^2+J_1^2}-J(1-\Delta)/2$).
The ground state exactly at a critical field can be built from any combination of singlet or polarized states on vertical bonds.
In the case we are going out of critical field the degeneracy disappears and the energy of state starts to depend on the states of neighboring bonds:
\begin{eqnarray}
\label{ham_spin}
H^{MFI-SB}=\sum_{i=1}^{N/2}\frac{(h-h_c^{(1/2)})}{4}(1-\sigma^z_i\sigma^z_{i+1}),
\end{eqnarray}
where $\sigma_i^z=\pm 1$ correspond to the polarized spin up and singlet bond states.

The free energy and all thermodynamic functions for the models (\ref{ham_dimer}) and (\ref{ham_spin})
can be easily calculated using the transfer-matrix method (see e.g. \cite{zhitomirsky2004,derzhko2006}).
Using the thermodynamic relation for the specific heat per site
$c=-T(\partial^2F/\partial T^2)$, one can show that the specific heat behaves as
$c\sim (h-h_c)^2/T^2$ close to the critical field.

Fig.\ref{fig_heat} displays the field dependence of specific heat at low temperatures,
which shows a rapid decline near critical field.
Such a behaviour can be explained using the notion of the effective lattice-gas models considered above.
The specific heat of the lattice-gas models (\ref{ham_dimer}) and (\ref{ham_spin})
has minimum at a critical field that is enclosed by additional maxima. Hence, it follows
that the lattice-gas models provide a good description of the low-temperature specific heat in a vicinity of critical fields.
The complete form of the specific heat can be recovered if we sum up contributions near all critical points,
however, this procedure may over-estimate the specific heat at higher temperatures with respect to the exact results.

In conclusion, we have examined the low-temperature behavior of the specific heat
for the Heisenberg-Ising orthogonal-dimer chain. We observed that it has minima at the critical fields.
These minima has been explained within the effective lattice gas models formulated
in the vicinity of each magnetization jump.

\acknowledgements
T.V. acknowledges the support of the National Scholarship Programme of the Slovak Republic. J.S. acknowledges the financial support under the grant VEGA 1/0234/12.

\end{document}